\documentclass[aps,pra,twocolumn,superscriptaddress,showpacs,floatfix,longbibliography]{revtex4-2}

\usepackage{amsmath,amssymb,amsfonts}
\usepackage{tikz}
\usetikzlibrary{calc,positioning,arrows.meta,decorations.markings,
                 patterns,backgrounds}
\usepackage{booktabs}
\usepackage[
  breaklinks=true,
  colorlinks=true,
  citecolor=blue,
  linkcolor=blue,
  urlcolor=blue
]{hyperref}
\usepackage{orcidlink}
\usepackage{xcolor}

\definecolor{ctxA}{RGB}{200,80,80}
\definecolor{ctxB}{RGB}{80,80,200}
\definecolor{ctxC}{RGB}{80,160,80}
\definecolor{ctxD}{RGB}{200,140,40}
\definecolor{ctxE}{RGB}{140,40,200}

\begin{document}

\title{Complementarity in Social Measurement: A Partition-Logic Approach}

\author{Karl Svozil\,\orcidlink{0000-0001-6554-2802}}
\affiliation{Institute for Theoretical Physics, TU Wien, Wiedner Hauptstrasse 8-10/136, A-1040 Vienna, Austria}
\email{karl.svozil@tuwien.ac.at}

\date{\today}

\begin{abstract}
Partition logics---non-Boolean event structures obtained by pasting Boolean algebras---provide a natural language for situations in which a system has a definite latent state but can be accessed and resolved only through mutually incompatible coarse-grained modes of observation. We show that this structure arises in a range of social-science settings by constructing six explicit examples from personnel assessment, survey framing, clinical diagnosis, espionage coordination, legal pluralism, and organizational auditing. For each case we identify the latent state space, the observational contexts as partitions, and the shared atoms that intertwine contexts, yielding instances of the $L_{12}$ bowtie, triangle, pentagon, and automaton partition logics. These examples make precise a notion of social complementarity: different modes of inquiry can be incompatible even though the underlying system remains fully value-definite. Complementarity in this sense does not entail contextuality or ontic indeterminacy. We further compare the classical probabilities generated by convex mixtures of dispersion-free states with the quantum-like Born probabilities available when the same exclusivity graph admits a faithful orthogonal representation. The framework thus separates logical structure from probabilistic realization and suggests empirically testable benchmarks for quantum-cognition models.
\end{abstract}

\maketitle

\section{Introduction}
\label{sec:intro}

Complementarity, the impossibility of simultaneously observing all
relevant properties of a system, is usually associated with quantum
mechanics. Niels Bohr famously argued that the wave and particle
aspects of a quantum object cannot be revealed in a single
experimental arrangement. Yet the \emph{formal} structure underlying
complementarity---a collection of Boolean ``classical mini-universes''
pasted together at shared elements---appears naturally in
many social-science measurement situations. In these situations the
system under study (an applicant, a survey respondent, an
organization) possesses fully determinate properties, but the
observer is forced to choose among incompatible modes of inquiry, each
of which reveals only a coarse-grained view of the true state.

The mathematical framework that captures this situation is the
\emph{partition logic}~\cite{svozil-93,schaller-92,
dvur-pul-svo,svozil-2001-eua}. A partition logic is built from a
finite set of underlying types $S_n = \{1,2,\ldots,n\}$.
Each ``context'' or ``block'' is a partition of $S_n$ into
groups; elements in the same group are those types that \emph{cannot
be distinguished} when that particular observational context is
chosen. Two contexts are then ``pasted'' together whenever they share
an identical group of types---an ``intertwining atom.'' The resulting
algebraic structure is, in general, non-Boolean: the join or meet of
propositions from different contexts need not be defined, reflecting
the fact that one cannot combine information from incompatible
observations in a straightforward logical way.

The purpose of this paper is to make this abstract formalism concrete
through six social-science scenarios that span organizational
psychology, political science, clinical psychology, security studies,
comparative law, and public administration. For each scenario we will
answer three questions explicitly:
\begin{enumerate}
\item What are the \textbf{types} (elements of $S_n$), and
what do they represent in the social world?
\item What are the \textbf{contexts} (partitions), and what
observational procedure does each one correspond to?
\item What are the \textbf{intertwining atoms} (shared partition
elements), and why does the same group of types happen to be
indistinguishable under two different procedures?
\end{enumerate}
We will also discuss a remarkable mathematical fact: on every such
partition logic, \emph{two} distinct probability theories can be
defined. The first is the ``natural'' classical one, in which
probabilities are convex combinations of dispersion-free ($0/1$) truth
assignments. The second is a ``quantum-like'' Born-rule probability
computed from a faithful orthogonal representation of the graph.
These two probability theories make different numerical predictions,
and the difference is in principle empirically testable---a fact of
considerable interest for the quantum-cognition
program~\cite{Busemeyer2012,Pothos2013}.
There may exist more than two probability theories satisfying Kolmogorov-type axioms on blocks (i.e., classical mini-universes);
in particular,
(i) exclusivity (probabilities of mutually exclusive events sum up),
(ii) completeness (probabilities of mutually exclusive events sum up to $1$).
For an explicit example, we shall encounter a third type when discussing pentagon/pentagram logics.

While the history and sociology of the quantum mechanics community are fascinating subjects in their own right~\cite{Howard2004-HOWWIT,Cabello-2015-madness}, an analysis of these aspects falls outside the scope of this paper.

Section~\ref{sec:prelim} reviews the formalism.
Sections~\ref{sec:personnel}--\ref{sec:org} present the six
scenarios. Section~\ref{sec:discussion} discusses the dual
probability interpretation. Section~\ref{sec:conclusion} concludes.

\section{Preliminaries: Partition logics and their probabilities}
\label{sec:prelim}

Before delving into the specific social-science scenarios, we first establish the formal mathematical framework of partition logics, their graphical representations, and the dual probability theories they support.
This provides a rigorous foundation for modeling complementarity in social measurements.

\subsection{The elements: latent states, observed categories, and pasting}

A partition logic starts from a finite set
$S = \{s_1, s_2, \ldots, s_n\}$ of \emph{latent states}.
Each latent state is one fully specified possibility in the model:
for example, a voter who is economically left-leaning and
culturally conservative, or a household that is above the poverty
line but materially deprived, or a job applicant whose true
competence profile is ``strong analytical skills, weak
interpersonal skills.'' We assume that, in reality, the system
under study \emph{is} exactly one latent state $s \in S$, even if
the observer cannot always determine which one. The substantive
interpretation of a latent state---what it means for a person,
household, student, or organization to be in state $s$---is
called the \emph{latent social profile} associated with $s$. The
word ``latent'' signals only that the full profile is not directly
read off from any single observational procedure; it does
\emph{not} imply that the profile is indeterminate or unreal.

Potentially, the set $S$ of latent states captures the
\emph{ontology} of the situation: what the system truly is.
By contrast, what can be \emph{inferred} about the system---and
what distinctions cannot be drawn with a particular instrument---belongs
to the \emph{epistemology}. That epistemological structure is
encoded by partitions of $S$.

A \emph{context} is a partition
$\mathcal{C} = \{B_1, B_2, \ldots, B_m\}$ of $S$ into $m$
nonempty, pairwise disjoint subsets whose union is $S$. Each
context represents one available mode of observation: a survey
item, an administrative classification rule, a diagnostic
instrument, a coding scheme, or a framing device. Each cell
$B_j \subseteq S$ of the partition is called an \emph{observed
category} (synonymously, an \emph{atom} in the language of
partition logic). An observed category is the coarsest unit of
information that the chosen context can deliver: when the observer
selects context $\mathcal{C}$, all latent states within the same
observed category $B_j$ produce indistinguishable outcomes.
The observer learns only \emph{which observed category} the true
latent state belongs to, not which specific latent state it is.
Thus the partition encodes the resolving power---and the
limitations---of a particular mode of inquiry.

To summarize the ontology--epistemology mapping:
\begin{itemize}
\item A \textbf{latent state} $s \in S$ is one exact underlying
  possibility (ontology).
\item A \textbf{latent social profile} is the substantive
  social-science interpretation of that latent state---what it
  means to be a Star applicant, a Cognitive-dominant patient, a
  financially sound organization, etc.\ (ontology, interpreted).
\item A \textbf{context} $\mathcal{C}$ is a partition of $S$
  induced by a particular instrument, survey frame, legal code,
  or audit directive (epistemology: choice of observational mode).
\item An \textbf{observed category} $B_j \subseteq S$ is one cell
  of that partition---the group of latent states that the chosen
  context cannot tell apart (epistemology: what is actually seen).
\end{itemize}

Two contexts $\mathcal{C}$ and $\mathcal{C}'$ may share an
observed category: a subset $B \subseteq S$ may appear as an atom
in both partitions.
(They may, in principle, share more than just one atom corresponding to more than just one latent state.)
This means that there is a group of latent
states that ``look the same'' regardless of which of the two
observational procedures is used. Such a shared observed category
is called an \emph{intertwining atom}. It is the formal point at
which the two Boolean subalgebras---each representing the
propositional logic of one context, in which all distinctions are
classical and simultaneously decidable---are ``pasted'' together
to form the full, non-Boolean partition
logic~\cite{greechie-1974,wright:pent,svozil-2018-b}.

\emph{Complementarity} manifests itself in the fact that different
contexts partition the latent-state space differently: observed
categories that are resolved (split into finer groups) by one
context may be conflated (merged into a single group) by another.
Crucially, every latent state retains its determinate identity
throughout---the complementarity is a limitation on the
\emph{observer's} ability to ascertain the latent social profile,
not an indeterminacy in the profile itself.

\subsection{Graphical representations}
\label{sec:prelim-gr}

Two equivalent graphical notations are common. In a (uniform) hypergraph~\cite{Bretto-MR3077516}
(often also referred to as a \emph{Greechie orthogonality diagram})~\cite{greechie:71}, each context is drawn as a smooth line and
each atom as a vertex on that line; intertwining atoms sit at the
intersection of two (or more) lines. In the \emph{exclusivity graph},
vertices are atoms and edges connect atoms that are mutually exclusive
(i.e., belong to the same context).

Uniform hypergraphs are characterized by an identical number of vertices on every hyperedge. In standard quantum logic, an $n$-dimensional Hilbert space naturally generates an $n$-uniform hypergraph, because the number of dimensions strictly dictates the number of elements in any orthonormal basis (and its corresponding set of orthogonal projection operators). In the social sciences, however, this strict uniformity is not required. Complementary observational procedures may yield maximal sets of mutually exclusive outcomes of varying sizes; formally stated, the blocks comprising the partition logic may be Boolean algebras of different orders~\cite{greechie:71}. While this marks a structural departure from standard quantum mechanics, it is not a decisive obstacle. It does not fundamentally alter the probabilistic framework, nor does it affect deeper logical properties such as embeddability into a global Boolean algebra (which corresponds to the existence of hidden variables). Because the concrete scenarios presented in this paper rely on straightforward uniform hypergraphs, a more extensive discussion of non-uniform structures falls outside the present scope and is relegated to future work.

\subsection{Multiple probability theories on the same structure}
\label{sec:twoprob}

Henceforth, any admissible probability measure must satisfy two Kolmogorov-type conditions on each block (i.e., classical mini-universe)~\cite{Gleason}:
\begin{enumerate}
\item  \textbf{Additivity/exclusivity}: the probability of a union of mutually exclusive events within the same block is the sum of their individual probabilities (cf.\ Cauchy's functional equation~\cite{Wright2019}); and
\item  \textbf{Normalization/completeness}: the probabilities of all mutually exclusive events comprising a block must sum to $1$.
\end{enumerate}

A \emph{dispersion-free state} (or two-valued weight) is a function
$v \colon \text{atoms} \to \{0,1\}$ that assigns~$1$ to exactly one atom
per context. Intuitively, it encodes the statement, ``the system \emph{is} this type
and no other.'' Every dispersion-free state corresponds to one of the
$n$ latent types in $S_n$.

\paragraph{Classical probabilities.}
Given a population of systems (applicants, respondents, etc.), the
fraction of each type defines a probability distribution over
$S_n$. The resulting probabilities on atoms are convex
combinations of the dispersion-free weights: $P(B_j) = \sum_{i \in
B_j} \lambda_i$, where $\lambda_i \geq 0$ and $\sum_i \lambda_i =
1$. The set of all such distributions forms the \emph{vertex packing
polytope} $\text{VP}(G)$ of the exclusivity
graph~\cite{GroetschelLovaszSchrijver1986}---a convex polytope whose vertices are
the dispersion-free states.

\paragraph{Quantum-like (Born-rule) probabilities.}
If the exclusivity graph $G$ admits a \emph{faithful orthogonal
representation} (FOR)---an assignment of unit vectors $|v_i\rangle$
in $\mathbb{R}^d$ to atoms such that vectors within the same context
are orthogonal and vectors from different contexts are
non-orthogonal~\cite{lovasz-79,Cabello-2014-gtatqc}---one can define a
``probability'' by choosing a unit ``state'' vector $|c\rangle$ and
setting
\begin{equation}
P_{\text{Born}}(c, v_i) = |\langle c | v_i \rangle|^2.
\label{eq:born}
\end{equation}
By the Pythagorean theorem, within each context the probabilities
add to one. The set of all distributions obtainable this way is the
\emph{theta body} $\text{TH}(G)$~\cite{GroetschelLovaszSchrijver1986}, which in
general \emph{strictly contains} $\text{VP}(G)$: there exist Born-rule
distributions that cannot be realized by any classical mixture of
dispersion-free weights.

Crucially, the same formal graph can
be implemented by a social-science ``black box'' (yielding classical
probabilities from $\text{VP}(G)$) \emph{or} by a quantum resource (yielding
Born-rule probabilities from $\text{TH}(G)$). The graph alone does
not determine the probability theory; the \emph{resource inside the
box} does~\cite{svozil-2018-b}.

See Section~\ref{2026-pliss:exoticProbs} for a discussion of another ``exotic'' type of probability measure, as exposed by Gerelle, Greechie, and Miller~\cite[Fig.~V]{greechie-1974} as well as Wright~\cite{wright:pent}.

\section{Scenario 1: Complementary personnel assessments ($L_{12}$)}
\label{sec:personnel}

In what follows, we develop several concrete instances and enactments of the formal scenario involving the \emph{partition logics} introduced earlier.


For related discussions and additional examples, we refer the reader to the extensive works of
Aerts and colleagues~\cite{aerts1995applications,aerts1995applications,aerts2005a,aerts2009experimental,aerts2013concepts,aerts2022human,aerts2023thermodynamics,aerts2024quantum,aerts2025quantum},
Khrennikov and colleagues~\cite{Khrennikov2010,Khrennikov2009-decision,Haven2013-qss,Khrennikov2015-modeling,Khrennikov2018-neuronal,khrennikov2020social,Khrennikov2020-rationality},
and numerous scholars~\cite{franco2009conjunction,lambert2009type,aerts2010quantum,yukalov2010mathematical,lambert2012quantum,atmanspacher2012order,yukalov2014conditions,Ashtiani2015-survey,wendt2015quantum,Asano2015-adaptivity,orrell2016quantum,yukalov2018quantitative,orrell2020bvalue,arioli2021quantum,Meghdadi2022-biases,holtfort2023social,Aspalter2024,Aspalter2024b,Aspalter2025}.
These authors have explored similar quantum-like structures across cognition, game theory~\cite{eisert1999quantum,sanzmartin2025mapping}, and decision-making.

\subsection{What are the types?}

A firm's human resources (HR) department faces a pool of job applicants. Each applicant
has a true, latent \emph{competence profile}, falling into one of
four categories:
\begin{itemize}
\item \textbf{Type~1 (``Star''):} Excellent both in analytical
reasoning and in interpersonal communication.
\item \textbf{Type~2 (``Analyst''):} Strong analytical skills but
weaker interpersonal skills.
\item \textbf{Type~3 (``Communicator''):} Strong interpersonal
skills but weaker analytical ability.
\item \textbf{Type~4 (``Developing''):} Still developing in both
dimensions.
\end{itemize}
Thus $S_4 = \{1,2,3,4\}$, and every applicant \emph{is}
exactly one type---the uncertainty is the HR department's, not the
applicant's.

\subsection{What are the contexts?}

Two assessment instruments are available:
\begin{itemize}
\item \textbf{Written aptitude test ($C_W$):} A timed analytical
exam. It can identify Stars (type~1) by their top scores and
Analysts (type~2) by their strong-but-not-top scores, but it cannot
distinguish Communicators from Developing applicants, because both
score poorly on a purely analytical task. The three observable
outcomes are therefore: ``Outstanding'' $=\{1\}$, ``Adequate'' $=
\{2\}$, ``At-risk'' $= \{3,4\}$.
\item \textbf{Behavioral interview ($C_I$):} A structured
interpersonal exercise. It can identify Stars (type~1) by their
social poise, and Communicators (type~3) by their good-but-not-top
social skills, but it cannot distinguish Analysts from Developing
applicants, because both perform weakly in interpersonal tasks. The
three outcomes are: ``Outstanding'' $= \{1\}$, ``Adequate'' $=
\{3\}$, ``At-risk'' $= \{2,4\}$.
\end{itemize}

\begin{table}[b]
\caption{\label{tab:personnel}Response profiles for four
applicant types under two complementary instruments. Each row is
an applicant type; each column shows how that type appears under
the given instrument.}
\begin{ruledtabular}
\begin{tabular}{lcc}
Applicant type & Written test outcome & Interview outcome \\
\colrule
1: ``Star'' & Outstanding & Outstanding \\
2: ``Analyst'' & Adequate & At-risk \\
3: ``Communicator'' & At-risk & Adequate \\
4: ``Developing'' & At-risk & At-risk \\
\end{tabular}
\end{ruledtabular}
\end{table}

Crucially, suppose the two instruments \textbf{cannot be administered
independently} to the same applicant. Administering the written test
first induces ``stereotype threat'' or test anxiety that contaminates
subsequent interview performance; conversely, the social priming of
an interview biases subsequent test-taking behavior (anchoring
effects~\cite{Tversky1974}). The HR department must therefore
\emph{choose one instrument per applicant}---this is the source of
complementarity.

Formally, the two contexts are represented by the partitions $\{\{1\},\{2\},\{3,4\}\}$
and $\{\{1\},\{3\},\{2,4\}\}$, or, in epistemic terms,
\begin{equation}
\begin{split}
&\{\text{Star},\text{Analyst},\text{Developing}\}  \\
&\text{ and } \\
& \{\text{Star},\text{Communicator}, \text{Developing}\}
.
\end{split}
\end{equation}

\subsection{What is the intertwining atom?}

The atom $\{1\}$ (``Outstanding'') appears identically in both
partitions (Table~\ref{tab:personnel}). This is because the Star
(type~1) excels on \emph{both} dimensions, so regardless of which
instrument is chosen, Stars are singled out. In the language of
partition logic, $\{1\}$ is the intertwining atom at which the two
three-element Boolean algebras $2^3$ are pasted together.

Note what is \emph{not} shared: the ``At-risk'' label has a
\emph{different meaning} in each context---$\{3,4\}$ under the
written test vs.\ $\{2,4\}$ under the interview. Although the label
is the same, the underlying sets of types are different, so these are
\emph{not} intertwining atoms. Only atoms that correspond to exactly
the same set of types in both contexts are identified.

\begin{figure}[t]
\centering
\begin{tikzpicture}[scale=0.90, transform shape, every path/.style={line width=1pt}]

\draw[red] (0,2)--(0.9,1)--(1.8,0);
\draw[blue]   (1.8,0)--(2.7,1)--(3.6,2);

\filldraw[fill=red, draw=red] (0,2) circle (3pt);
\node[left,yshift=-10pt] at (0,2) {$\{2\}=\text{Analyst}$};

\filldraw[fill=red, draw=red] (0.9,1) circle (3pt);
\node[left,yshift=-10pt] at (0.9,1) {$\{3,4\}=\text{Developing}$};

\filldraw[fill=red, draw=red] (1.8,0) circle (4pt);
\filldraw[fill=blue,   draw=blue]   (1.8,0) circle (2pt);
\node[below,yshift=-10pt] at (1.8,0) {$\{1\}=\text{Star}$};

\filldraw[fill=blue, draw=blue] (2.7,1) circle (3pt);
\node[right,yshift=-10pt] at (2.7,1) {$\{2,4\}=\text{Developing}$};

\filldraw[fill=blue, draw=blue] (3.6,2) circle (3pt);
\node[right,yshift=-10pt] at (3.6,2) {$\{3\}=\text{Communicator}$};

\end{tikzpicture}
\caption{Greechie orthogonality diagram (hypergraph) of the
$L_{12}$ personnel-assessment logic. Each line represents one
context---one available mode of observation. The
{\color{ctxA}written-test context} $\mathcal{C}_W$ (left) carries
three observed categories: the singleton $\{2\}$ (Analyst,
recognizable by adequate test performance), the pair
$\{3,4\}$ (Communicator and Developing, conflated because
both score poorly on the analytical task), and the singleton
$\{1\}$ (Star, top scorer). The {\color{ctxB}interview context}
$\mathcal{C}_I$ (right) carries a different triple of observed
categories: $\{3\}$ is now resolved (Communicator, identifiable
by social poise) but $\{2,4\}$ is conflated (Analyst and
Developing, both weak in interpersonal tasks). The observed
category $\{1\}$ is the \emph{intertwining atom}: it
appears in both contexts because the Star's latent social
profile---excellence on both dimensions---makes this profile
identifiable regardless of which instrument is chosen. The five
observed categories, together with their lattice-theoretic joins
and meets, generate a $12$-element non-Boolean logic, hence the
name~$L_{12}$.}
\label{fig:L12}
\end{figure}
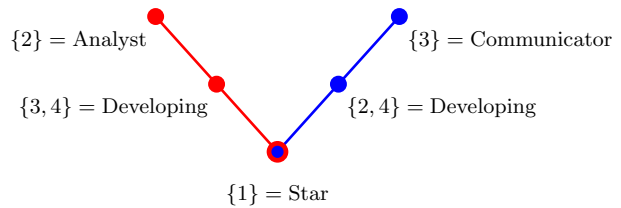

Formally, the two contexts, depicted in Fig.~\ref{fig:L12}, are thus represented by the partitions $\{\{1\},\{2\},\{3,4\}\}$
and $\{\{1\},\{3\},\{2,4\}\}$, or, in epistemic terms,
\begin{equation}
\begin{split}
&\{\text{Star},\text{Analyst},\text{Developing}\}  \\
&\text{ and } \\
& \{\text{Star},\text{Communicator}, \text{Developing}\}
.
\end{split}
\end{equation}

\subsection{What does the partition logic tell us?}

The partition logic $L_{12}$ (Fig.~\ref{fig:L12}) encodes a precise
tradeoff. By choosing the written test, the HR department gains the
ability to distinguish type~2 from types~3 and~4, at the cost of
being unable to distinguish type~3 from type~4. By choosing the
interview, it can distinguish type~3 from types~2 and~4, but now
type~2 and type~4 are conflated. No single instrument resolves all
four types. This is \emph{complementarity}: the information about
one aspect of the applicant (analytical vs.\ interpersonal) comes at
the expense of information about the other. Yet every applicant
\emph{has} a determinate type; the partition logic captures an
epistemic limitation of the observer, not an ontic indeterminacy of
the system.

\section{Scenario 2: Framing effects in public opinion surveys
(Pentagon logic)}
\label{sec:surveys}

Our second scenario turns to political psychology, demonstrating how well-documented question-order and framing effects in public opinion surveys naturally generate the cyclic structure of a pentagon partition logic.

\subsection{What are the types?}

A political psychologist studies a population whose members differ
in their latent configuration of attitudes across five policy
domains. Rather than continuous attitude scales, suppose (for
simplicity) that the relevant attitudinal differences can be captured
by $n=11$ discrete respondent \emph{profiles}, denoted
$S_{11}=\{1,2,\ldots,11\}$. Each profile specifies a
particular pattern of beliefs about immigration, healthcare,
taxation, education, and defense spending. For example:
\begin{itemize}
\item Profile~1 might be a ``consistent moderate'' who holds centrist
views on all five issues.
\item Profile~7 might be a ``libertarian'' who favors minimal
government across all domains.
\item Profile~10 might be a ``hawk-dove'' who supports high defense
spending as well as generous social programs.
\end{itemize}
Suppose the 11 profiles are exhaustive and mutually exclusive: every
respondent \emph{is} exactly one profile.

\subsection{What are the contexts?}

The five policy issues define five \emph{survey contexts}. In each
context, the respondent is asked a block of questions about one issue
(e.g., immigration), and their response is classified into one of
three \emph{opinion clusters}: (A) supportive, (B) moderate, (C)
opposed (or similar trichotomy). Crucially, due to well-documented
\textbf{question-order and framing effects}~\cite{Moore2002}, the
way respondents categorize their own position on an issue is
influenced by the issue discussed immediately before. In
survey-methodology terms, asking about immigration ``primes''
concepts (e.g., national identity, fiscal cost) that alter responses
to a subsequent healthcare question.

For this reason, in any single session a respondent can be
meaningfully polled on only \emph{one} issue---attempting to cover
all five in sequence would yield contaminated data that does not
reflect the respondent's true underlying profile. This forced choice
of survey frame is the source of complementarity.

The five contexts, adapted from the partition logic of the
pentagon~\cite{greechie-1974,wright:pent,svozil-2018-b}, are:
\begin{align}
C_1\;\text{(immigration)} &= \bigl\{\underbrace{\{1,2,3\}}_{\text{Cluster
A}},\;\underbrace{\{4,5,7,9,11\}}_{\text{Cluster
B}},\;\underbrace{\{6,8,10\}}_{\text{Cluster C}}\bigr\},
\notag\\
C_2\;\text{(healthcare)} &= \bigl\{\{6,8,10\},\;\{1,2,4,7,11\},\;
\{3,5,9\}\bigr\},
\notag\\
C_3\;\text{(taxation)} &= \bigl\{\{3,5,9\},\;\{1,4,6,10,11\},\;
\{2,7,8\}\bigr\},
\label{eq:pentagon}\\
C_4\;\text{(education)} &= \bigl\{\{2,7,8\},\;\{1,3,9,10,11\},\;
\{4,5,6\}\bigr\},
\notag\\
C_5\;\text{(defense)} &= \bigl\{\{4,5,6\},\;\{7,8,9,10,11\},\;
\{1,2,3\}\bigr\}.
\notag
\end{align}
Each context partitions all 11 profiles into three clusters.

\subsection{What are the intertwining atoms?}

Consider contexts $C_1$ (immigration) and $C_2$ (healthcare). The
atom $\{6,8,10\}$ appears as Cluster~C in $C_1$ and as Cluster~A in
$C_2$. Substantively, this means that profiles~6, 8, and~10 are
respondents who (i)~oppose the prevailing immigration policy
\emph{and} (ii)~support the prevailing healthcare policy---and
\emph{both of these facts are observable regardless of whether the
survey asks about immigration or healthcare}. The cluster
$\{6,8,10\}$ is an intertwining atom precisely because these three
profiles happen to ``look the same'' under both frames.

\begin{figure}[t]
\centering
\begin{tikzpicture}  [scale=0.3]

\newdimen\ms
\ms=0.05cm

\tikzstyle{every path}=[line width=1pt]

\tikzstyle{c3}=[circle,inner sep={\ms/8},minimum size=6*\ms]
\tikzstyle{c2}=[circle,inner sep={\ms/8},minimum size=3*\ms]
\tikzstyle{c1}=[circle,inner sep={\ms/8},minimum size=0.8*\ms]

\newdimen\R
\R=6cm     

\path
  ({90 + 0 * 360 /5}:\R      ) coordinate(1)
  ({90 + 36 + 0 * 360 /5}:{\R * sqrt((25+10*sqrt(5))/(50+10*sqrt(5)))}      ) coordinate(2)
  ({90 + 1 * 360 /5}:\R   ) coordinate(3)
  ({90 + 36 + 1 * 360 /5}:{\R * sqrt((25+10*sqrt(5))/(50+10*sqrt(5)))}   ) coordinate(4)
  ({90 + 2 * 360 /5}:\R  ) coordinate(5)
  ({90 + 36 + 2 * 360 /5}:{\R * sqrt((25+10*sqrt(5))/(50+10*sqrt(5)))}  ) coordinate(6)
  ({90 + 3 * 360 /5}:\R  ) coordinate(7)
  ({90 + 36 + 3 * 360 /5}:{\R * sqrt((25+10*sqrt(5))/(50+10*sqrt(5)))}  ) coordinate(8)
  ({90 + 4 * 360 /5}:\R     ) coordinate(9)
  ({90 + 36 + 4 * 360 /5}:{\R * sqrt((25+10*sqrt(5))/(50+10*sqrt(5)))}     ) coordinate(10)
;


\draw [color=orange] (1) -- (2) -- (3);
\draw [color=red] (3) -- (4) -- (5);
\draw[color=green] (5) -- (6) -- (7);
\draw [color=blue] (7) -- (8) -- (9);
\draw [color=magenta] (9) -- (10) -- (1);    %

%
%
\draw (1) coordinate[c3,fill=orange,label=90:{\footnotesize $\{ 1,2,3\} $}];   %
\draw (1) coordinate[c2,fill=magenta];  %
\draw (2) coordinate[c3,fill=orange,label={above left:\footnotesize $\{ 7,8,9,10,11\}$}];    %
\draw (3) coordinate[c3,fill=red,label={left:\footnotesize $\{ 4,5,6\} $}]; %
\draw (3) coordinate[c2,fill=orange];  %
\draw (4) coordinate[c3,fill=red,label={left:\footnotesize $\{ 1,3,9,10,11\}$}];  %
\draw (5) coordinate[c3,fill=green,label={left:\footnotesize $\{ 2,7,8\} $}];  %
\draw (5) coordinate[c2,fill=red];  %
\draw (6) coordinate[c3,fill=green,label={below:\footnotesize $\{ 1,4,6,10,11\} $}];
\draw (7) coordinate[c3,fill=blue,label={right:\footnotesize $\{ 3,5,9\}$}];  %
\draw (7) coordinate[c2,fill=green];  %
\draw (8) coordinate[c3,fill=blue,label={right:\footnotesize $\{ 1,2,4,7,11\}$}];  %
\draw (9) coordinate[c3,fill=magenta,label={right:\footnotesize $\{ 6,8,10\}$}];
\draw (9) coordinate[c2,fill=blue];  %
\draw (10) coordinate[c3,fill=magenta,label={above right:\footnotesize $\{ 4,5,7,9,11\}$}];  %
\end{tikzpicture}
\caption{Pentagon survey logic. Each color corresponds to a context
(policy issue)
corresponding to
stances on immigration, healthcare, taxation, education, and defense spending (clockwise from top right). The cyclic structure
reflects the pattern of framing overlaps among the five issues.}
\label{fig:pentagon}
\end{figure}
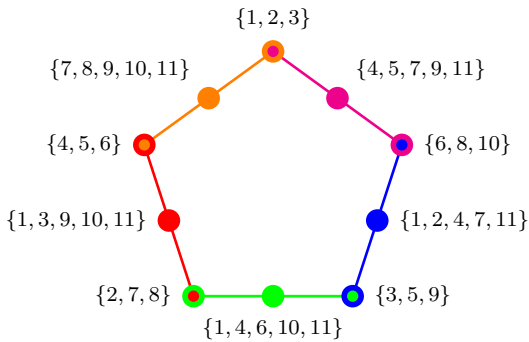

The five intertwining atoms, one per adjacent pair of contexts,
form the outer vertices of the pentagon diagram
(Fig.~\ref{fig:pentagon}). They are:
$\{1,2,3\}$ (shared by $C_1$ and $C_5$),
$\{6,8,10\}$ (shared by $C_1$ and $C_2$),
$\{3,5,9\}$ (shared by $C_2$ and $C_3$),
$\{2,7,8\}$ (shared by $C_3$ and $C_4$),
$\{4,5,6\}$ (shared by $C_4$ and $C_5$).

Formally, the five cyclically intertwined contexts (or cliques) form a pentagon/pentagram logic~\cite[p.~267, Fig.~2]{wright:pent}
supporting $11$ dispersion-free states. Constructing the five contexts from the occurrences of the dispersion-free value $1$ on the respective ten atoms
results in the partition logic:
\begin{equation}
\begin{aligned}
&\{
 \{ \{ 1,2,3\} ,\{ 4,5,7,9,11\} ,\{ 6,8,10\} \} ,\\
&\{ \{ 6,8,10\} ,\{ 1,2,4,7,11\} ,\{ 3,5,9\} \} ,\\
&\{ \{ 3,5,9\} ,\{ 1,4,6,10,11\} ,\{ 2,7,8\} \} ,\\
&\{ \{ 2,7,8\} ,\{ 1,3,9,10,11\} ,\{ 4,5,6\} \} ,\\
&\{ \{ 4,5,6\} ,\{ 7,8,9,10,11\} ,\{ 1,2,3\} \}
\},
\end{aligned}
\label{2018-e-plpentagon}
\end{equation}
as depicted in Fig.~\ref{fig:pentagon}.

\subsection{Why the pentagon structure matters: classical vs.\
quantum-like bounds}

Each intertwining atom is a subset of the $11$ profiles. The
\emph{classical} probability that a randomly drawn respondent falls
in any one of these subsets is simply the fraction of the population
belonging to those profiles. The key constraint is that the five
intertwining-atom subsets overlap: for instance, profile~3 belongs
to both $\{1,2,3\}$ and $\{3,5,9\}$. Classical counting shows that
each of the $11$ profiles can contribute to at most two of the five
intertwining atoms, so
\begin{equation}
\sum_{j=1}^{5} P(\text{intertwining atom}_j) \;\leq\; 2.
\label{eq:classical_bound}
\end{equation}
This is the Klyachko--Bub--Stairs
inequality~\cite{Klyachko-2008,Bub-2009}. It says: no matter how
the population is distributed across the $11$ profiles, the total
``intertwining-atom weight'' cannot exceed~2.

If, however, the respondents' cognitive states are modeled as
vectors in a Hilbert space---as proposed in the quantum-cognition
literature~\cite{Busemeyer2012}---then probabilities are
$|\langle c|v_j\rangle|^2$ and can be tuned (by choosing the state
$|c\rangle$) to yield
\begin{equation}
\sum_{j=1}^{5} P(\text{intertwining atom}_j) \;\leq\; \sqrt{5}
\approx 2.236.
\label{eq:quantum_bound}
\end{equation}
This $12\%$ increase over the classical bound is an empirically
testable prediction. A survey experiment that (i)~implements the
pentagon exclusivity structure and (ii)~finds aggregate
intertwining-atom probabilities exceeding~2 would constitute
evidence against a classical partition-logic model and in favor of a
quantum-like cognitive model.

\section{Scenario 3: Diagnostic complementarity in clinical
psychology (triangle logic)}
\label{sec:clinical}

In our third scenario, we turn to clinical diagnostics, demonstrating how cyclic interference between different psychological and physiological assessment tools naturally gives rise to a triangle partition logic.

\subsection{What are the latent states?}

A clinical psychologist suspects that patients presenting with
generalized anxiety disorder (GAD) actually fall into four latent
states---four diagnostic subtypes that differ along two
dimensions (cognitive worry and physiological arousal) but are
never fully revealed by any single instrument:
\begin{itemize}
\item \textbf{Latent state $1$ (``Cognitive-somatic''):} The
  patient experiences both high cognitive worry (rumination,
  catastrophizing) \emph{and} high physiological arousal
  (elevated cortisol, reduced heart-rate variability). This is
  the most floridly symptomatic subtype.
\item \textbf{Latent state $2$ (``Cognitive-dominant''):} High
  cognitive worry but low physiological arousal. The patient
  ruminates intensely but does not show the somatic signature of
  anxiety.
\item \textbf{Latent state $3$ (``Somatic-dominant''):} Low
  cognitive worry but high physiological arousal. The patient
  reports feeling ``fine'' yet shows marked autonomic
  dysregulation.
\item \textbf{Latent state $4$ (``Subclinical''):} Moderate
  levels of both worry and arousal that fall below typical
  clinical thresholds on \emph{either} dimension taken alone.
  This subtype is the hardest to classify, because it never
  produces a salient signal on any single instrument.
\end{itemize}
The set of latent states is thus
$S=\{1,2,3,4\}$. Each patient \emph{is} exactly one
subtype; the diagnostic challenge is that no single instrument
can identify all four.

\subsection{What are the three contexts?}

Three diagnostic instruments are available, each resolving two
of the three ``clear'' subtypes ($1$, $2$, $3$) as
singleton observed categories while \emph{conflating} the third
with the elusive $4$:

\begin{itemize}
\item \textbf{Context $\mathcal{C}_A$ --- Self-report
  questionnaire (cognitive focus).} The patient completes a
  standardized inventory of worry-related cognitions. Instrument~A
  picks up the cognitive dimension:
  \begin{itemize}
  \item $1$ (Cognitive-somatic) reports extreme worry $\;\to\;$
    observed category $\{1\}$ (``severe cognitive'').
  \item $2$ (Cognitive-dominant) reports high worry $\;\to\;$
    observed category $\{2\}$ (``moderate cognitive'').
  \item $3$ and $4$ both report little cognitive worry
    $\;\to\;$ conflated observed category $\{3,4\}$
    (``low cognitive'').
  \end{itemize}
  Partition:
  $\mathcal{C}_A = \bigl\{\{1\},\;\{2\},\;\{3,4\}\bigr\}$.

\item \textbf{Context $\mathcal{C}_B$ --- Behavioral observation
  (interpersonal focus).} A clinician trained in standardized behavioral coding observes the patient during a structured social-interaction task (e.g., a simulated conversation) and systematically rates observable signs such as gaze aversion, fidgeting, tremor, and speech latency.
  Instrument~B picks up the \emph{behavioral expression} of
  anxiety:
  \begin{itemize}
  \item $2$ (Cognitive-dominant) displays visible avoidance,
    gaze aversion, and social withdrawal driven by ruminative
    worry $\;\to\;$ observed category $\{2\}$ (``avoidant'').
  \item $3$ (Somatic-dominant) displays visible tremor,
    fidgeting, and perspiration from physiological arousal
    $\;\to\;$ observed category $\{3\}$ (``tremorous'').
  \item $1$ and $4$ are conflated: $1$ (Cognitive-somatic)
    has learned to compensate socially despite high symptom load,
    and $4$ (Subclinical) simply does not display salient signs
    $\;\to\;$ conflated observed category $\{1,4\}$
    (``unremarkable presentation'').
  \end{itemize}
  Partition:
  $\mathcal{C}_B = \bigl\{\{2\},\;\{3\},\;\{1,4\}\bigr\}$.

\item \textbf{Context $\mathcal{C}_C$ --- Physiological stress
  test (somatic focus).} The patient undergoes a standardized
  stress-induction protocol while heart-rate variability and
  cortisol are recorded. Instrument~C picks up the somatic
  dimension:
  \begin{itemize}
  \item $1$ (Cognitive-somatic) shows extreme autonomic
    reactivity $\;\to\;$ observed category $\{1\}$
    (``severe physiological'').
  \item $3$ (Somatic-dominant) shows elevated reactivity
    $\;\to\;$ observed category $\{3\}$ (``moderate
    physiological'').
  \item $2$ and $4$ both show unremarkable physiological
    profiles $\;\to\;$ conflated observed category
    $\{2,4\}$ (``low physiological'').
  \end{itemize}
  Partition:
  $\mathcal{C}_C = \bigl\{\{1\},\;\{3\},\;\{2,4\}\bigr\}$.
\end{itemize}

\begin{table*}[htb]
\caption{\label{tab:clinical}How the four latent diagnostic
subtypes map to observed categories under each of the three
instruments. Each row is a latent state; each column shows the
observed category it falls into under the given context. Boldface
marks singleton observed categories (the instrument
\emph{resolves} that subtype); plain text marks the conflated
pair (the instrument \emph{cannot} distinguish the two subtypes
in that group).}
\begin{ruledtabular}
\begin{tabular}{lccc}
Latent state & $\mathcal{C}_A$ (Self-report) &
  $\mathcal{C}_B$ (Behavioral) & $\mathcal{C}_C$ (Physiological)\\
\colrule
$1$ (Cognitive-somatic)  & \textbf{severe cognitive}  & unremarkable
  & \textbf{severe physiological}\\
$2$ (Cognitive-dominant)  & \textbf{moderate cognitive} & \textbf{avoidant}
  & low physiological\\
$3$ (Somatic-dominant)  & low cognitive  & \textbf{tremorous}
  & \textbf{moderate physiological}\\
$4$ (Subclinical) & low cognitive  & unremarkable
  & low physiological\\
\end{tabular}\end{ruledtabular}
\end{table*}

Table~\ref{tab:clinical} summarizes the mapping. The crucial
pattern is that $4$ (Subclinical) \emph{never} forms a
singleton: under every instrument it is conflated with a
different partner---with $3$ under~A, with $1$ under~B, and
with $2$ under~C. It is the ``stealth'' subtype that each
instrument misses in a different way.

\subsection{Why only one instrument per session?}

The three instruments interfere with one another in a cyclic
pattern that prevents sequential administration within a single
diagnostic session:
\begin{enumerate}
\item \emph{A degrades B.} Completing the self-report
  questionnaire induces introspective self-focus, which alters
  the patient's spontaneous behavior in the subsequent social
  interaction~\cite{Duval1972}. The behavioral observation
  would then reflect questionnaire-primed comportment rather
  than the patient's natural presentation.
\item \emph{B degrades C.} The social interaction of being
  observed by a rater elevates the patient's baseline
  physiological arousal (social-evaluative
  threat~\cite{Dickerson2004}), contaminating the subsequent
  stress-test readings.
\item \emph{C degrades A.} Undergoing a physiological
  stress-induction protocol triggers mood-congruent recall and
  somatic focusing~\cite{Bower1981}, biasing the patient's
  subsequent self-reports toward overendorsement of cognitive
  worry.
\end{enumerate}
These interferences are not merely practical inconveniences;
they are systematic disturbances that change the patient's
momentary state in ways relevant to the other instruments. The
clinician must therefore \emph{choose exactly one instrument per
session}---this forced choice is the source of complementarity.

\subsection{What are the intertwining atoms?}

Each pair of contexts shares exactly one observed category
(Table~\ref{tab:intertwine_clinical}):
\begin{itemize}
\item $\{1\}$ is an intertwining atom of $\mathcal{C}_A$ and
  $\mathcal{C}_C$: both the self-report questionnaire and the
  physiological stress test single out the Cognitive-somatic
  subtype, because $1$'s extreme scores on both the cognitive
  \emph{and} the somatic dimension make it identifiable
  regardless of whether one probes cognition or physiology.
\item $\{2\}$ is an intertwining atom of $\mathcal{C}_A$ and
  $\mathcal{C}_B$: the questionnaire flags $2$ as ``moderate
  cognitive'' and the behavioral observation flags $2$ as
  ``avoidant''---different labels, but the \emph{same set} of
  latent states, $\{2\}$. The Cognitive-dominant subtype is
  identifiable whether one reads the patient's self-report or
  watches the patient's behavior.
\item $\{3\}$ is an intertwining atom of $\mathcal{C}_B$ and
  $\mathcal{C}_C$: the behavioral observation sees tremor and the
  stress test sees elevated cortisol---both pointing uniquely to
  the Somatic-dominant subtype.
\end{itemize}

\begin{table*}[htb]
\caption{\label{tab:intertwine_clinical}Intertwining atoms of the
triangle logic. Each row names a pair of contexts, the shared
observed category, and the clinical reason why the subtype is
identifiable under both instruments.}
\begin{ruledtabular}
\begin{tabular}{lll}
Context pair & Shared atom & Reason for identifiability\\
\colrule
$\mathcal{C}_A$--$\mathcal{C}_C$ & $\{1\}$ &
  Extreme on \emph{both} worry and arousal\\
$\mathcal{C}_A$--$\mathcal{C}_B$ & $\{2\}$ &
  High worry shows in report \emph{and} behavior\\
$\mathcal{C}_B$--$\mathcal{C}_C$ & $\{3\}$ &
  High arousal shows in behavior \emph{and} physiology\\
\end{tabular}
\end{ruledtabular}
\end{table*}

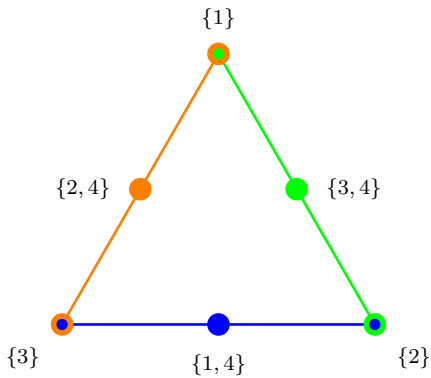
\begin{figure}[t]
\centering
\begin{tikzpicture}  [scale=0.4]

\newdimen\ms
\ms=0.05cm

\tikzstyle{c3}=[circle,inner sep={\ms/8},minimum size=6*\ms]
\tikzstyle{c2}=[circle,inner sep={\ms/8},minimum size=3*\ms]
\tikzstyle{c1}=[circle,inner sep={\ms/8},minimum size=0.8*\ms]

\tikzstyle{every path}=[line width=1pt]

\newdimen\R
\R=6cm     

\path
  ({90 + 0 * 360 /3}:\R      ) coordinate(1)
  ({90 + 1 * 360 /3}:\R   ) coordinate(3)
  ({90 + 2 * 360 /3}:\R  ) coordinate(5)
;

\draw [color=orange] (1) -- (3);
\draw [color=blue] (3) -- (5);
\draw [color=green] (5) -- (1);

%
%
\draw (1) coordinate[c3,fill=orange];   %
\draw (1) coordinate[c2,fill=green,label=90:\colorbox{white}{\footnotesize  $\{1\}$}];  %
\node[c3,fill=orange,label={left:\colorbox{white}{\footnotesize   $\{2,4\}$}}] at ( $ (1)!{1/2}!(3) $ ) (2) {};    %
\draw (3) coordinate[c3,fill=orange];  %
\draw (3) coordinate[c2,fill=blue,label=below left:{\footnotesize  \colorbox{white}{$\{3\}$}}];  %
\node[c3,fill=blue,label={below:{\footnotesize  \colorbox{white}{$\{1,4\}$}}}] at ( $ (3)!{1/2}!(5) $ ) (4) {};  %
\draw (5) coordinate[c3,fill=green];  %
\draw (5) coordinate[c2,fill=blue,label=below right:{\footnotesize   \colorbox{white}{$\{2\}$}}];  %
\node[c3,fill=green,label={right:\colorbox{white}{\footnotesize  $\{3,4\}$}}] at ( $ (5)!{1/2}!(1) $ ) (6) {};  %
\end{tikzpicture}
\caption{Greechie diagram of the triangle logic applied
to clinical diagnosis. Each edge represents one context
(diagnostic instrument). The three yellow vertices are the
intertwining atoms---the singleton observed categories, each
identifiable under two adjacent instruments. The three white
vertices are the conflated observed categories, each merging
$4$ (Subclinical) with a different partner. $4$ is the
``stealth'' subtype: it never appears as a singleton under any
instrument, yet it has a fully determinate latent social profile.
The six observed categories, together with their
lattice-theoretic operations, form a non-Boolean logic that
cannot be embedded in any single Boolean algebra.}
\label{fig:triangle}
\end{figure}

These three intertwining atoms are the vertices of a triangle in
the Greechie diagram (Fig.~\ref{fig:triangle}). Each edge of the
triangle represents one context; the non-shared observed
categories (the conflated pairs $\{3,4\}$, $\{1,4\}$,
$\{2,4\}$) sit at the midpoints of the edges. The result is
Wright's \emph{triangle logic}~\cite[Figure~2, p.~900]{wright}
(see also~\cite[Fig.~6, Example~8.2, pp.~414,420,421]{dvur-pul-svo}).

\subsection{The exotic probability weight: clinical interpretation}
\label{2026-pliss:exoticProbs}

Wright~\cite{wright:pent} (see also Ref.~\cite[Fig.~V]{greechie-1974}) showed that cyclic logics with an odd number $n$ of intertwining atoms---including the triangle, pentagon, and pentagram logics---support an ``exotic'' probability weight in addition to their standard dispersion-free weights. This weight assigns a value of $1/2$ to each intertwining vertex and lies strictly \emph{outside} the classical convex hull, rendering this $\{0,1/2\}$-valued state inexpressible as a convex mixture of the $\{0,1\}$-valued dispersion-free states.

To see this, consider the sum of the weights assigned to the $n$ intertwining atoms, $S(\omega) = \sum_{i=1}^n \omega(v_i)$. Because these atoms are arranged in a cycle where each adjacent pair $\{v_i, v_{i+1}\}$ (with $v_{n+1}=v_1$) belongs to a common block, any valid weight must satisfy the consistency constraint $\omega(v_i) + \omega(v_{i+1}) \leq 1$. For the exotic state $\omega_{ex}$, the functional yields $S(\omega_{ex}) = n/2$. However, for any dispersion-free state $\omega_c \in \{0, 1\}$, the constraint forbids adjacent atoms from both being $1$. On an odd cycle, the maximum number of non-adjacent vertices that can be assigned $1$ is $(n-1)/2$; thus, $S(\omega_c) \leq (n-1)/2$. Since any state $\omega_{hull}$ in the classical convex hull is a mixture of these states, it is bounded by $S(\omega_{hull}) \leq (n-1)/2$. Because $n/2 > (n-1)/2$ for all odd $n$, the exotic state is strictly non-classical.

In support of the reality of such a scenario, Wright gave the following snippet~\cite[p.~272]{wright:pent}:
\begin{quote}
``\dots suppose that there is a consumer who
has no strong opinions one way or the other about the products, so that
he is equally likely to say either yes or no to the first question, but
suppose, further, that he is embarrassed about saying no~\cite{smith1975} so that
if he says no to the first question, then he always says yes to the
second.''
\end{quote}

What could this mean in the clinical context? It would represent a probability
distribution in which there is exactly a $50\%$ chance of
observing the identifiable subtype ($1$, $2$, or $3$) under
any given instrument---but this distribution is
\emph{inconsistent with any population mixture of the four
diagnostic subtypes}. No matter how one adjusts the prevalence
rates of the four subtypes, one cannot reproduce the exotic
weight by classical mixing.

If a clinician's empirical frequencies across a patient population
approached this exotic weight, it would suggest that the classical
partition-logic model (determinate subtypes, purely epistemic
uncertainty) is inadequate. One possible interpretation, drawn
from the quantum-cognition literature~\cite{Busemeyer2012}, is
that patients' diagnostic profiles are not fixed prior to
measurement but are instead \emph{constituted in part by the
diagnostic act itself}---the instrument does not merely reveal a
pre-existing subtype but participates in producing the
categorization. Whether this interpretation is warranted or
whether alternative explanations (e.g., systematic violations of
the single-instrument-per-session constraint) are more
parsimonious is an empirical question that the partition-logic
framework makes precise.

\section{Scenario 4: Espionage and relational encoding (classical
EPR analogue)}
\label{sec:espionage}

Our fourth scenario shifts from individual assessments to multiparty correlations. By modeling two spatially separated espionage agents sharing a predefined codebook, we construct a classical analogue to the Einstein-Podolsky-Rosen (EPR) setup.

\subsection{What are the types?}

An intelligence agency trains agent pairs using a protocol that
creates correlations between their cover stories. Each agent pair is
prepared in one of four \emph{coordination types}:
\begin{itemize}
\item \textbf{Type~1 (``Mirror''):} Both agents memorize identical
financial records and identical personal histories. (Code: 00---
same digit in both positions.)
\item \textbf{Type~2 (``Split-A''):} Agent~A memorizes a clean
financial record but a fabricated personal history; Agent~B
memorizes the reverse. (Code: 01.)
\item \textbf{Type~3 (``Split-B''):} The reverse of Split-A.
(Code: 10.)
\item \textbf{Type~4 (``Contrast''):} Both agents memorize
fabricated financial records and fabricated personal histories, but
with opposite details. (Code: 11.)
\end{itemize}

\subsection{What are the contexts?}

Upon capture, each agent faces one of two interrogation modes:
\begin{itemize}
\item \textbf{Financial background check (F):} The interrogator
examines only financial records. Agent's response is the first digit
of the code ($0 = \text{clean}$, $1 = \text{fabricated}$).
\item \textbf{Personal history interview (H):} The interrogator
examines only personal history. Agent's response is the second
digit of the code.
\end{itemize}
An agent subjected to one mode gives answers that are psychologically
``activated'' in a way that would contaminate a subsequent
interrogation of a different mode (a well-known carryover effect in
interrogation science). Thus only one mode can be applied per agent.

\subsection{Relational encoding and why it is local}

The key point is that the agency loads agent pairs from the
subensemble $ D  = \{01, 10\}$ (Table~\ref{tab:relational}),
meaning only types~2 and~3 are used. The relational encoding
guarantees that \textbf{if both agents face the same interrogation
mode, their answers are always opposite}: when Alice's financial
check yields ``clean'' ($0$), Bob's yields ``fabricated'' ($1$), and
vice versa. This perfect anticorrelation arises entirely from the
shared codebook---a \emph{classical common cause} established before
separation.

\begin{table}[b]
\caption{\label{tab:relational}Relational encoding of cover
stories for agent pairs.
Each entry is a two-digit code: first digit = financial record,
second digit = personal history.
$ S $ selects types whose codes match pairwise; $ D $ selects
types whose codes differ pairwise.}
\begin{ruledtabular}
\begin{tabular}{lcccc}
Sample & Type 1 & Type 2 & Type 3 & Type 4\\
\colrule
$ S $ (same)  & 00 &    &    & 11 \\
$ D $ (diff.) &    & 01 & 10 &    \\
\end{tabular}
\end{ruledtabular}
\end{table}

The partition logic structure is as follows. For each agent,
the two interrogation modes define two partitions of the type
space. For Alice:
\begin{align}
C_F^{(A)} &= \bigl\{\{1,2\},\;\{3,4\}\bigr\}
\quad\text{(first digit)}, \notag\\
C_H^{(A)} &= \bigl\{\{1,3\},\;\{2,4\}\bigr\}
\quad\text{(second digit)}.
\label{eq:alice}
\end{align}
For Bob, the same. Each agent's logic is the pasting of two
two-element Boolean algebras $2^2$---a minimal complementary
structure.

\subsection{Social science significance}

This scenario models any situation in which
two spatially or temporally separated actors produce correlated
behaviors through a \emph{shared preparation} rather than direct
communication:
\begin{itemize}
\item \textbf{Coordinated deception:} As described---espionage
cover stories.
\item \textbf{Institutional scripts:} Employees trained in the same
corporate protocol independently produce consistent behaviors when
facing customers~\cite{Cyert1963}.
\item \textbf{Shared cultural schemas:} Individuals socialized in
the same culture give correlated survey responses about values and
norms, even when separated by continents, because they internalized
the same ``codebook'' during childhood~\cite{bordieu-fu}.
\end{itemize}
In all cases, the correlations are \emph{local} (no signaling) and
\emph{classical} (probabilities lie within $\text{VP}(G)$). The scenario
demonstrates that correlation without communication is not uniquely
quantum; what \emph{is} uniquely quantum is the ability to
\emph{violate Bell inequalities}, which the shared-codebook
mechanism can never do.

\section{Scenario 5: Legal pluralism and overlapping jurisdictions}
\label{sec:legal}

Our fifth scenario applies the framework of partition logics to comparative law, illustrating how the existence of multiple, overlapping legal frameworks creates a formal structure of complementarity.

\subsection{What are the types?}

In a legally pluralistic society---for example, a post-colonial
state where statutory law, customary (indigenous) law, and religious
law coexist~\cite{Merry1988,Griffiths1986}---a citizen's action can
be classified differently depending on which legal framework is
applied. Consider six types of action involving land use:
\begin{itemize}
\item $a_1$: Clearing forest for subsistence farming on ancestral
land.
\item $a_2$: Clearing forest for commercial logging on ancestral
land.
\item $a_3$: Building a house on communally held village land.
\item $a_4$: Building a commercial structure on communally held
village land.
\item $a_5$: Grazing livestock on a protected nature reserve.
\item $a_6$: Grazing livestock on privately owned pasture.
\end{itemize}
Each action is a determinate fact; the question is how it is
\emph{categorized} by different legal codes.

\subsection{What are the contexts?}

Three legal codes, each classifying actions into three categories
(``Permitted,'' ``Restricted,'' ``Prohibited''), define three
contexts:
\begin{itemize}
\item \textbf{Statutory (environmental) law ($C_{\text{stat}}$):} Focuses
on environmental impact. Actions $a_1, a_2$ (forest clearing) are
\emph{Prohibited}; $a_3, a_4$ (building on non-forest land) are
\emph{Restricted} (requiring permits); $a_5, a_6$ (grazing) are
\emph{Permitted}. Partition: $\{\{a_1,a_2\}, \{a_3,a_4\},
\{a_5,a_6\}\}$.
\item \textbf{Customary (indigenous) law ($C_{\text{cust}}$):} Focuses on
ancestral rights and communal obligations. $a_1, a_3$ (subsistence
use of ancestral/communal land) are \emph{Permitted} (rights of
use); $a_2, a_5$ (commercial exploitation of communal resources)
are \emph{Restricted} (requiring elder consent); $a_4, a_6$ (use
of land outside the communal system) are outside customary
jurisdiction, categorized as \emph{Neutral}. Partition:
$\{\{a_1,a_3\}, \{a_2,a_5\}, \{a_4,a_6\}\}$.
\item \textbf{Religious law ($C_{\text{rel}}$):} Focuses on stewardship
obligations. $a_1, a_2$ (forest clearing) are \emph{Prohibited}
(the forest is sacred); $a_3, a_5$ (non-destructive use) are
\emph{Permitted}; $a_4, a_6$ (commercial use) are
\emph{Discouraged}. Partition: $\{\{a_1,a_2\}, \{a_3,a_5\},
\{a_4,a_6\}\}$.
\end{itemize}

\begin{table}[b]
\caption{\label{tab:legal}Classification of six land-use actions
under three legal codes. Actions in the same cell of a given
column are indistinguishable (conflated) under that code.}
\begin{ruledtabular}
\begin{tabular}{lccc}
Action & Statutory & Customary & Religious \\
\colrule
$a_1$ (subsist.\ clearing) & Prohibited & Permitted & Prohibited \\
$a_2$ (comm.\ logging)     & Prohibited & Restricted & Prohibited \\
$a_3$ (house on village)   & Restricted & Permitted & Permitted \\
$a_4$ (shop on village)    & Restricted & Neutral & Discouraged \\
$a_5$ (graze on reserve)   & Permitted  & Restricted & Permitted \\
$a_6$ (graze on private)   & Permitted  & Neutral & Discouraged \\
\end{tabular}
\end{ruledtabular}
\end{table}

\subsection{What are the intertwining atoms?}

Two atoms are shared:
\begin{itemize}
\item $\{a_1,a_2\}$ (``Prohibited'') is an intertwining atom of
$C_{\text{stat}}$ and $C_{\text{rel}}$: both statutory and religious law agree
that all forms of forest clearing are prohibited, regardless of
purpose. This agreement arises from a convergence of environmental
and spiritual rationales.
\item $\{a_4,a_6\}$ is an intertwining atom of $C_{\text{cust}}$ and
$C_{\text{rel}}$: both customary and religious law place commercial
land use in a separate category (``Neutral''/``Discouraged''),
treating it as outside the domain of traditional
obligations---albeit for different reasons (customary law sees it
as irrelevant; religious law sees it as morally suspect).
\end{itemize}

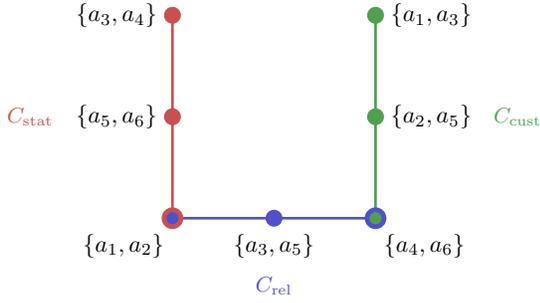
\begin{figure}[t]
\centering
\begin{tikzpicture}[scale=0.90, transform shape, every path/.style={line width=1pt}]

\tikzstyle{every path}=[line width=1pt]

\definecolor{statutory}{RGB}{200,80,80}    
\definecolor{customary}{RGB}{80,160,80}    
\definecolor{religious}{RGB}{80,80,200}    

\draw[statutory] (0,3) -- (0,1.5) -- (0,0);

\draw[religious] (0,0) -- (1.5,0) -- (3,0);

\draw[customary] (3,0) -- (3,1.5) -- (3,3);

\filldraw[fill=statutory, draw=statutory] (0,3) circle (3pt);
\node[left, font=\normalsize] at (-0.1,3) {$\{a_3, a_4\}$};

\filldraw[fill=statutory, draw=statutory] (0,1.5) circle (3pt);
\node[left, font=\normalsize] at (-0.1,1.5) {$\{a_5, a_6\}$};

\filldraw[fill=statutory, draw=statutory] (0,0) circle (4pt);
\filldraw[fill=religious, draw=religious] (0,0) circle (2pt);
\node[below left, font=\normalsize] at (0,-0.1) {$\{a_1, a_2\}$};

\filldraw[fill=religious, draw=religious] (1.5,0) circle (3pt);
\node[below, font=\normalsize] at (1.5,-0.1) {$\{a_3, a_5\}$};

\filldraw[fill=religious, draw=religious] (3,0) circle (4pt);
\filldraw[fill=customary, draw=customary] (3,0) circle (2pt);
\node[below right, font=\normalsize] at (3,-0.1) {$\{a_4, a_6\}$};

\filldraw[fill=customary, draw=customary] (3,1.5) circle (3pt);
\node[right, font=\normalsize] at (3.1,1.5) {$\{a_2, a_5\}$};

\filldraw[fill=customary, draw=customary] (3,3) circle (3pt);
\node[right, font=\normalsize] at (3.1,3) {$\{a_1, a_3\}$};

\node[statutory, font=\small] at (-2.1,1.5) {$C_{\text{stat}}$};
\node[religious, font=\small] at (1.5,-1) {$C_{\text{rel}}$};
\node[customary, font=\small] at (5.1,1.5) {$C_{\text{cust}}$};

\end{tikzpicture}
\caption{Hypergraph of the legal-pluralism logic. Three
legal codes (Statutory, Religious, Customary) form three contexts.
Yellow vertices are intertwining atoms: $\{a_1,a_2\}$ (forest
clearing---prohibited under both statutory and religious law) and
$\{a_4,a_6\}$ (commercial use---set apart by both customary and
religious law). White vertices represent groupings specific to one
code. The non-Boolean pasting reflects the impossibility of a
single ``master code'' that respects all three systems'
distinctions simultaneously.}
\label{fig:legal}
\end{figure}

\subsection{What does the non-Boolean structure mean for legal
practice?}

In practice, a land-use dispute is adjudicated under \emph{one}
legal framework at a time---the choice of forum (statutory court,
customary tribunal, or religious council) determines which
distinctions are drawn and which actions are conflated:
\begin{itemize}
\item Under statutory law, subsistence clearing ($a_1$) and
commercial logging ($a_2$) are treated identically (both
prohibited), whereas customary law treats them very
differently (permitted vs.\ restricted).
\item Under customary law, a house ($a_3$) and a shop ($a_4$) on
village land are in different categories, but under statutory law
both are merely ``restricted.''
\end{itemize}
The non-Boolean logic (Fig.~\ref{fig:legal}) formalizes the
well-known problem in comparative law that there is no single
``master code'' that simultaneously preserves all the distinctions
made by each individual legal system. The choice of forum is
analogous to the choice of measurement context in quantum
mechanics---and it is a genuine choice with real consequences for the
outcome.

\section{Scenario 6: Organizational auditing and bounded
rationality (automaton partition logic)}
\label{sec:org}

Our final scenario models organizational auditing, where the act of inspection inherently disturbs the corporation's internal state. This creates an automaton partition logic that formally captures Herbert Simon's concept of bounded rationality.

\subsection{What are the types?}

A government regulatory agency must assess whether a corporation
complies with regulations. The corporation can be in one of four
internal \emph{organizational states}, characterized by two binary
dimensions:
\begin{itemize}
\item \textbf{Financial health:} Sound ($0$) vs.\ Distressed ($1$).
\item \textbf{Operational compliance:} Compliant ($0$) vs.\
Non-compliant ($1$).
\end{itemize}
The four states are:
\begin{itemize}
\item $s_1 = (0,0)$: Financially sound \emph{and} operationally
compliant---the ideal.
\item $s_2 = (0,1)$: Financially sound but operationally
non-compliant---cutting corners despite having resources.
\item $s_3 = (1,0)$: Financially distressed but operationally
compliant---struggling but playing by the rules.
\item $s_4 = (1,1)$: Financially distressed \emph{and}
non-compliant---the worst case.
\end{itemize}
Each corporation \emph{is} exactly one of these states at any given
time.

\subsection{What are the contexts?}

The auditor can probe the corporation by issuing one of two types
of directive:
\begin{itemize}
\item \textbf{Financial audit (directive $a$):} The auditor
requests detailed financial records. The corporation's response
reveals its \emph{financial health} (the first digit), but
the act of preparing financial disclosures causes the organization
to reallocate resources internally, thereby \emph{changing its
operational compliance}. (Employees are pulled off production lines
to prepare audit documents, maintenance is deferred, etc.) The
auditor observes output~0 (sound) or~1 (distressed), but the
original compliance state is destroyed.
Partition: $C_a = \{\{s_1,s_2\},\{s_3,s_4\}\}$.
\item \textbf{Operational inspection (directive $b$):} The auditor
sends inspectors to the factory floor. The response reveals
\emph{operational compliance} (the second digit), but
the disruption of an on-site inspection causes financial
perturbations (e.g., halted production, legal fees). The original
financial state is destroyed.
Partition: $C_b = \{\{s_1,s_3\},\{s_2,s_4\}\}$.
\end{itemize}

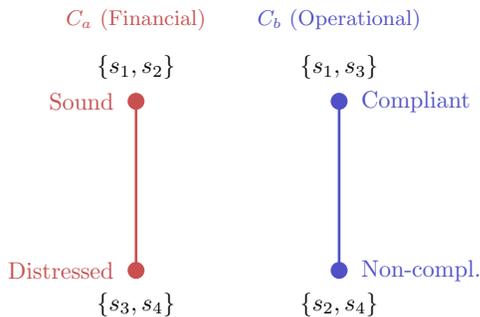
\begin{figure}[t]
\centering
\begin{tikzpicture}[scale=0.90, transform shape, every path/.style={line width=1pt}]

\tikzstyle{every path}=[line width=1pt]

\definecolor{financial}{RGB}{200,80,80}     
\definecolor{operational}{RGB}{80,80,200}   

\draw[financial] (0,0) -- (0,2.5);

\filldraw[fill=financial, draw=financial] (0,2.5) circle (3pt);
\node[above, font=\normalsize] at (0,2.7) {$\{s_1, s_2\}$};
\node[left, xshift=-10pt, font=\normalsize, financial] at (0.15,2.5) {Sound};

\filldraw[fill=financial, draw=financial] (0,0) circle (3pt);
\node[below, font=\normalsize] at (0,-0.2) {$\{s_3, s_4\}$};
\node[left, xshift=-10pt, font=\normalsize, financial] at (0.15,0) {Distressed};

\draw[operational] (3,0) -- (3,2.5);

\filldraw[fill=operational, draw=operational] (3,2.5) circle (3pt);
\node[above, font=\normalsize] at (3,2.7) {$\{s_1, s_3\}$};
\node[right, xshift=10pt, font=\normalsize, operational] at (2.85,2.5) {Compliant};

\filldraw[fill=operational, draw=operational] (3,0) circle (3pt);
\node[below, font=\normalsize] at (3,-0.2) {$\{s_2, s_4\}$};
\node[right, xshift=10pt, font=\normalsize, operational] at (2.85,0) {Non-compl.};

\node[financial, font=\small] at (0,3.7) {$C_a$ (Financial)};
\node[operational, font=\small] at (3,3.7) {$C_b$ (Operational)};

\end{tikzpicture}
\caption{The four organizational states arranged in a $2\times 2$
grid. The financial audit (directive~$a$, {\color{ctxA}red
dashed}) partitions the grid horizontally: it reveals the row
(financial health) but conflates columns (compliance status).
The operational inspection (directive~$b$, {\color{ctxB}blue
dotted}) partitions vertically: it reveals the column but
conflates rows. The two partitions share \emph{no} atoms---this is
maximal complementarity.}
\label{fig:org_grid}
\end{figure}

\subsection{What are the intertwining atoms?}

In this scenario, the two partitions share \emph{no} intertwining
atoms (Fig.~\ref{fig:org_grid}):
\begin{align}
C_a &= \bigl\{\underbrace{\{s_1,s_2\}}_{\text{Sound}},\;
\underbrace{\{s_3,s_4\}}_{\text{Distressed}}\bigr\},
\notag\\
C_b &= \bigl\{\underbrace{\{s_1,s_3\}}_{\text{Compliant}},\;
\underbrace{\{s_2,s_4\}}_{\text{Non-compl.}}\bigr\}.
\label{eq:org}
\end{align}
The set $\{s_1,s_2\}$ (``Sound'') is not the same as $\{s_1,s_3\}$
(``Compliant'')---no atom appears in both contexts. This means there
is no organizational property that can be ascertained regardless of
audit type; \emph{every} piece of information the auditor obtains
is context-dependent in the sense that it tells the auditor about
one dimension while destroying information about the other. This
is \emph{complete complementarity}---stronger than the $L_{12}$
case, where at least one type (the Star) was identifiable under
either instrument.

\subsection{Connection to bounded rationality}

This formalization connects directly to Herbert Simon's concept of
\emph{bounded rationality}~\cite{Simon1957}: an external regulator
(or market analyst, or journalist) attempting to understand an
organization faces an irreducible limitation. The organization's
internal state is fully determinate, but no single probe can reveal
it completely, and probing inevitably disturbs the system.

The partition logic captures this as a structural feature, not
merely a practical inconvenience. The non-Boolean pasting of the two
two-element Boolean algebras means that the auditor's total
knowledge \emph{cannot be represented as a single probability
distribution over a single Boolean event space}. This is a formal
expression of organizational opacity---the well-known difficulty of
assessing complex organizations from the
outside~\cite{Cyert1963}.

The automaton aspect adds a further layer: issuing directive~$a$
causes the organization to \emph{transition to a new state},
meaning that a subsequent probe with~$b$ reveals information about
the \emph{post-audit} state, not the original one. This is
precisely the measurement-disturbance problem of quantum mechanics,
realized here by a completely classical mechanism: the audit
process itself consumes resources and alters the system.

\section{Discussion: Dual probabilities and quantum cognition}
\label{sec:discussion}

Having established the ubiquity of partition logics across six diverse social-science domains, we now explore a crucial mathematical feature of these structures: their ability to simultaneously support multiple, distinct probability theories.

\subsection{Summary of the dual interpretation}

All six scenarios share a common formal skeleton: a partition logic
pasted from Boolean subalgebras over a finite type set
$S_n$. On each such logic, two or more
probability theories may coexist:

\begin{enumerate}
\item \textbf{Classical probabilities} from the convex hull of
dispersion-free weights ($\text{VP}(G)$). These are the ``natural''
probabilities for the social science scenarios as described: every
applicant, respondent, patient, agent, action, or organization
\emph{is} a determinate type, and uncertainty is purely epistemic
(arising from the observer's inability to probe all contexts
simultaneously). Classical probabilities obey the
Klyachko--Bub--Stairs inequality~\eqref{eq:classical_bound} for the
pentagon and analogous inequalities for other graphs.

\item \textbf{Quantum-like (Born-rule) probabilities} from a
faithful orthogonal representation ($\text{TH}(G)$). These
probabilities can exceed the classical bounds---for the pentagon, up
to $\sqrt{5}$ instead of~2. They would be appropriate if the
``resource inside the black box'' were a quantum system rather than
a classical partition logic.
\end{enumerate}

\subsection{Why this matters for social science}

\paragraph{The resource determines the probability theory.}
The same exclusivity graph (e.g., the pentagon) can be implemented
by a Wright urn (yielding $\text{VP}(G)$) or by a quantum system (yielding
$\mathrm{TH}(G)$). Similarly, a survey with pentagon structure
could yield different probability bounds depending on whether the
respondents' cognitive processes are ``classical'' (definite beliefs
selected by a coarse-grained observation) or ``quantum-like''
(superposed belief states projected by the survey question).
The graph alone does not settle the matter; the \emph{nature of the
cognitive resource} does.

\paragraph{Complementarity does not imply contextuality.}
All six scenarios feature complementarity (the inability to
perform all measurements simultaneously and consistently; that is, with context-independent valuations~\cite{kochen1}), yet every element has a
determinate value in every dispersion-free state. The partition
logics have a separating set of two-valued states and are therefore
\emph{not} contextual in the Kochen--Specker
sense~\cite{kochen1}. In the social-science analogues, this is
intuitive: the applicant has a real competence profile, the patient
has a real subtype, the organization is in a real state---the
complementarity is epistemic rather than ontic---it is the observer's problem, not the system's.

\paragraph{Testable predictions.}
The dual-probability framework suggests concrete experiments.
Design a social-measurement scenario with a known exclusivity
structure (e.g., five cyclically complementary survey questions).
Collect data from many respondents, each answering only one
question. Compute the empirical probabilities on the intertwining
atoms. If the sum exceeds the classical bound~($2$ for the pentagon),
the classical partition-logic model is falsified, and a quantum-like
model is supported. This program extends the work of Busemeyer and
Bruza~\cite{Busemeyer2012} from pairwise order effects to the
richer graph-theoretic structures of partition logics.

\paragraph{Ontological, irreducible uncertainty?}
So far, we have presumed that uncertainty and complementarity arise because the underlying system properties are latent and our observational means are limited. This constitutes a purely epistemic stance, positing a pre-existing, determinate ontology that remains partially hidden from the observer. But what if these properties are not merely latent, but irreducibly indeterminate~\cite{zeil-05_nature_ofQuantum}---in theological terms, \textit{creatio continua}? A related, potentially weaker assumption---one that does not require the total abandonment of causality and the principle of sufficient reason---is that certain properties actively emerge from, or are instantiated by, the act of measurement itself. To model such behavior, one would have to go beyond partition logics and adopt more radical formal concepts, such as Kochen--Specker-type contextuality~\cite{kochen1}, or the emergence of macro-irreversibility from micro-reversibility (e.g., via infinite limits or the inflow of environmental information).

\paragraph{Non-uniform hypergraphs and varying block sizes.}
As noted in Section~\ref{sec:prelim-gr},
the six scenarios presented here all happen to produce uniform
hypergraphs: every context yields the same number of mutually
exclusive observed categories. This uniformity mirrors the
situation in standard quantum logic, where an $n$-dimensional
Hilbert space forces every orthonormal basis---and hence every
context---to contain exactly $n$~elements. In the social sciences,
however, there is no \emph{a priori} reason for such regularity.
Different observational procedures may well resolve the latent-state
space into maximal sets of mutually exclusive outcomes of unequal
size; formally, the blocks comprising the partition logic may be
Boolean algebras of different orders~\cite{greechie:71}.
A job interview that distinguishes three outcome categories and a
psychometric test that distinguishes five would generate a
non-uniform hypergraph. Crucially, this structural departure from
standard quantum mechanics is not a fundamental obstacle: the
Kolmogorov-type (admissability) axioms of additivity and normalization
(Section~\ref{sec:twoprob}) apply block by block and remain
well-defined regardless of whether all blocks have the same
cardinality. Likewise, deeper logical properties---most importantly,
the existence or non-existence of a separating set of two-valued
states, which governs embeddability into a global Boolean algebra
and thereby the viability of a hidden-variable model---are unaffected
by non-uniformity. Because all concrete examples in this paper are
uniform, a systematic exploration of non-uniform partition logics in
social-science measurement---including the modifications required for
faithful orthogonal representations and the associated probability
polytopes---is deferred to future work.

\section{Conclusion}
\label{sec:conclusion}

We have demonstrated that partition logics---non-Boolean structures
originally developed in the foundations of quantum mechanics, generalized urn models, and
automata theory~\cite{svozil-2001-eua}---arise naturally in six distinct social-science
domains. In each case, a finite set of underlying types (applicant
profiles, respondent attitudes, diagnostic subtypes, agent
coordination types, legal actions, organizational states) is
partitioned by multiple incompatible observational procedures
(assessment instruments, survey frames, diagnostic tests,
interrogation modes, legal codes, audit directives). The resulting
structures exhibit complementarity: no single procedure reveals all
relevant distinctions. Yet every system has a fully determinate
type; the limitation is epistemic, not ontic.

The social-science scenarios therefore occupy a kind of ``purgatory'' between classical
Boolean and quantum realms: they have non-Boolean logic and
complementarity, yet retain full value definiteness~\cite{svozil-2018-b}. This
demonstrates concretely that \emph{complementarity does not imply
contextuality}---a point often obscured in discussions that focus
exclusively on quantum mechanics.

At the same time, the mathematical fact that many partition logics
admit a dual, quantum-like probability interpretation (via faithful
orthogonal representations and the Born rule) opens a door to
empirical testing. If human cognitive or social systems turn out to
produce probability distributions that violate the classical bounds
but respect, say, the quantum-like or more exotic ones, the partition-logic framework
provides the precise mathematical language in which to formulate
and test this hypothesis. Conversely, if the classical bounds are
respected, the partition logic with its convex hull of
dispersion-free weights provides a complete and parsimonious account
of the observed complementarity---no quantum formalism required.

The development of experiments that probe these questions---for
instance, large-scale surveys with carefully designed pentagon
structures, or clinical assessment batteries with triangle-logic
configurations---is a promising direction for future work at the
intersection of formal logic, quantum foundations, and social
science methodology.

\begin{acknowledgments}
This research was funded in whole or in part by the \textit{Austrian Science Fund (FWF)}[Grant \textit{DOI:10.55776/PIN5424624}].
The authors acknowledge TU Wien Bibliothek for financial support through its Open Access Funding Programme.
\end{acknowledgments}

\bibliography{svozil}

\end{document}